\documentclass[conference]{IEEEtran}
\IEEEoverridecommandlockouts

\usepackage[T2A,T1]{fontenc}
\usepackage[utf8]{inputenc}
\usepackage[russian,english]{babel}
\usepackage{epsfig,endnotes}
\usepackage{comment}
\usepackage{hyperref}

\ifdefined\final
 \newcommand{\authorsnote}[2]{}
 \else
 \newcommand*{\authorsnote}[2]{\textcolor{#1}{[#2]}}
\fi

\providecommand{\hf}{Hack Forums}

\author{\IEEEauthorblockN{
Rasika Bhalerao\IEEEauthorrefmark{1},
Maxwell Aliapoulios\IEEEauthorrefmark{1}, 
Ilia Shumailov\IEEEauthorrefmark{2},
Sadia Afroz\IEEEauthorrefmark{3}, Damon McCoy\IEEEauthorrefmark{1}}
\IEEEauthorblockA{
\IEEEauthorrefmark{1} \textit{New York University},
\IEEEauthorrefmark{2} \textit{University of Cambridge},
\IEEEauthorrefmark{3} \textit{International Computer Science Institute},
\\
rasikabh@nyu.edu, maliapoulios@nyu.edu, Ilia.Shumailov@cl.cam.ac.uk, sadia@icsi.berkeley.edu, mccoy@nyu.edu,
}
}

\usepackage[table]{xcolor}
\usepackage{booktabs}
\usepackage{subcaption}
\usepackage{graphicx}
\graphicspath{{images/}}
\usepackage[linesnumbered,ruled,vlined]{algorithm2e}
\usepackage{algorithmic}
\usepackage{framed}
\usepackage{cite}

\def\BibTeX{{\rm B\kern-.05em{\sc i\kern-.025em b}\kern-.08em
    T\kern-.1667em\lower.7ex\hbox{E}\kern-.125emX}}
    
\title{Towards Automatic Discovery of Cybercrime Supply Chains}

\begin{document}

\maketitle

\begin{abstract}

Cybercrime forums enable modern criminal entrepreneurs to collaborate with other criminals into increasingly efficient and sophisticated criminal endeavors. 
Understanding the connections between different products and services can often illuminate  effective interventions. However, generating this understanding of supply chains currently requires time-consuming manual effort.

In this paper, we propose a language-agnostic method to automatically extract supply chains from cybercrime forum posts and replies. Our supply chain detection algorithm can identify 36\% and 58\% relevant chains within major English and Russian forums, respectively, showing improvements over the baselines of 13\% and 36\%, respectively. Our analysis of the automatically generated supply chains demonstrates underlying connections between products and services within these forums. For example, the extracted supply chain illuminated the connection between hack-for-hire services and the selling of rare and valuable `OG' accounts, which has only recently been reported. The understanding of connections between products and services exposes potentially effective intervention points.

\end{abstract}


\section{Introduction}
\label{introduction}
Cybercrime-as-a-Service lowers the barrier to entry for new cybercriminals by commoditizing different parts of attacks~\cite{43798}. For example, without commoditization, a spammer needs to find a way to send e-mails, acquire mailing lists, create storefront websites, contract with web hosting, register domains, manage product fulfillment, accept online payments, and provide customer service. With commoditization, the spammer can outsource different responsibilities to different criminals specialized in one specific task. Cybercriminals often rely on underground cybercrime forums to establish these trade relationships that can facilitate the exchange of illicit goods and services. 
These cybercrime forums thus play a crucial role in increasing efficiency and promoting innovation in the cybercrime ecosystem. However, these dependencies can also provide opportunities to undermine a spammer at more vulnerable points such as the online payment channel~\cite{Levchenko:2011}.

The supply chain of a cybercrime can illuminate the sequence of processes involved in the criminal activities. Prior work has shown that analyzing these supply chains can result in identifying weak points which could enable effective interventions~\cite{Clayton:WEIS15}.
There have been several studies exploring specific instance of these commoditized cybercrime offerings~\cite{Caballero:2011:MPC:2028067.2028080,Christin:2010:DOC:1866307.1866310,9e8f6c1fa27b4c83a3bccc54ab451c0c} and how some attacks can be more effectively undermined once their dependencies to other services are understood~\cite{182939,Karami:2016:STB:2872427.2883004,10.1007/978-3-642-32946-3_4,McCoy:2012:PRP:2382196.2382285}. 
However, we as a community do not have any systematic methods of identifying these supply chains that enable more sophisticated and streamlined attacks. Currently analysts often manually investigate cybercrime forums to understand these supply chains, which is a time-consuming process~\cite{krebsPC}.


In this paper, we propose, implement, and evaluate a framework to systematically identify relevant supply chains present in cybercrime forums. Our framework is composed of several components which include automated methods based on supervised Natural Language Processing (NLP) techniques and a graph-traversal algorithm. 
Our approach classifies the product category from a forum post, identifies the replies indicating that a user bought or sold a product, then builds an interaction graph and uses a graph traversal algorithm to discover links of related product buying and subsequent selling posts.
Our approach is language agnostic and does not require manual categorizations of products, an improvement over prior work on product detection~\cite{portnoff2017tools}.

We used our end-to-end supply chain identification pipeline to analyze two publicly available cybercrime forums and are able to identify 36\% and 58\% relevant links in our English language and Russian language forums, respectively. These are supply chain links where users are buying products that are likely used to facilitate subsequent product offerings (i.e., a user buying OSN reputation boosting services to groom accounts that are then sold to scammers) or users reselling products after they are no longer useful to their original owner. This is an increase from our baselines of 13\% and 36\%, respectively.

The main contributions of our paper are the following:
\begin{enumerate}
    \item[*] We develop, implement and evaluate an automated approach for discovering the cybercrime supply chain (Section~\ref{approach}). Our method uses language agnostic NLP methods and graph traversal to automate the discovery of the supply chains.
    
    \item[*] We perform an analysis of our automatically generated supply chains to provide an understanding of how some commodity cybercrime products depend on other offerings within these forums (Section~\ref{analysis}).
    
    \item[*] We distill our findings from the detected supply chains into several qualitative case studies (Section~\ref{case_studies}). These case studies highlight that we were efficiently able to discover supplies chains exposing the connection between the purchasing of hack-for-hire services and the selling of valuable online accounts. Despite this connection being present in the forums for years, it has only recently been discovered based on manual analysis~\cite{krebsswap}. This connection suggests a potentially more effective method of mitigating the theft of valuable online accounts by avoiding account authentication and recovery methods based on mobile phone numbers.

\end{enumerate}

The rest of this paper is structured in the following way. Section \ref{related} discusses background and related work in this area, including past work which uses the same data. Section \ref{data} outlines the data used to validate our approach. Section \ref{approach} outlines our approach and contributions in classification and supply chain discovery.  
We evaluate our work empirically  to demonstrate how our approach performs better than the baseline in Section \ref{eval}, and analyze the forums using our results in Section~\ref{analysis}. Section \ref{case_studies} outlines several real world scenarios where the generated supply chains add value to an investigation. Section~\ref{discussion} identifies limitations and discusses the implication of our results. We conclude in Section~\ref{conclusion}. 

\begin{table*}[!ht]
\centering
\begin{tabular}{@{}llllll@{}}
\toprule
                    & \multicolumn{1}{c}{\textbf{Total threads}} & \multicolumn{1}{c}{\textbf{Total replies}} & \multicolumn{1}{c}{\textbf{Unique authors}} & \multicolumn{1}{c}{\textbf{Total messages}} & \multicolumn{1}{c}{\textbf{Date range}} \\ \midrule
\textbf{Antichat}   & 73,115                                     & 287,089                                    & 20,540                                      & 328,215                                     & 05/2003 - 06/2010                       \\
\textbf{\hf} & 14,447                                     & 46,786                                     & 12,625                                      & 61,233                                      & 04/2009 - 04/2015                       \\ \bottomrule
\end{tabular}
\caption{Forum overviews}
\label{table:descriptive_statistics}
\end{table*}

\section{Background and Related Work}
\label{related}



Cybercriminal forums provide a unique opportunity to understand how criminal markets operate. Criminals rely on these forums to establish trade relationships and facilitate exchanges of goods and information.
The typical forum structure follows the \texttt{subforum} $>$ \texttt{thread} $>$ \texttt{post} $>$ \texttt{reply} hierarchy. A subforum typically pertains to a particular subject: for example, \texttt{marketplace} or \texttt{introductions}. Users then can create threads within subforums where a thread is a collection of messages. Each thread will always contain a first post, which in this paper, we will use interchangeably with \texttt{product post}. We will also use the term \texttt{reply post} to refer to the posts which come after the product post in each thread.  This is a key part of detecting relevant supply chains based on public information.





Several prior works studied the organization of the cybercrime forums~\cite{lusthaus2013organised,leukfeldt2016cybercriminal,motoyama2011analysis,abbasi2014descriptive}, profiling key actors~\cite{pastrana2018characterizing}, products traded~\cite{soska2015measuring,yip2012,Franklin2007,Holt2010}, evolution over time~\cite{allodi2016then,holt2013examining}, and ways to disrupt their business~\cite{huang2017cybercrime,decary2015discrediting}. However, these works either rely on the structural information on a forum or use handcrafted regular expression. 
Some prior work used machine learning to scale the analysis of these forums.

Portnoff et al.~\cite{portnoff2017tools} used supervised machine learning to automatically identify the type of a post (buy or sell), products being traded and the price of the products. Unlike their approach, our approach is language agnostic, which we demonstrate by analyzing both English and Russian forums. Caines et al.~\cite{Caines2018} recently explored classifying posts by intents, which is similar to how we classify replies as indicating buying or selling activity. Unlike their approach to classifying replies, we focus on identifying replies that strongly indicate that the forum member has actually bought or sold that product as opposed to expressing an intent to buy or sell a product. This is more useful for detecting likely supply chains.

Other work has explored the progression of illicit activities by forum members~\cite{pastrana2018crimebb}. Wegberg et al.~\cite{VW+:USENIXSec18} analyzed longitudinal data from eight structured online anonymous marketplaces over six years to understand the value change and commoditization of the criminal markets as ``cybercrime-as-a-service’’. They concluded that commoditization in the cybercrime market is still limited. 
Our work complements and extends this line of research by providing a method for detecting connections between products. 


Our approach goes beyond understanding the trust establishment, organization, aggregate activity, and classification performed in prior work. We use the results of our classifiers to identify semantically meaningful forum interactions and automatically discover supply chains of the products that can improve our understanding of how these markets function in practice. We analyzed the connections between products in unstructured cybercrime forums and noticed mostly business-to-business (we scope this to mean ``sale to the trade'' where the products being bought and sold often have no value except as a building block to enable an attack) transactions. This allows us to study the criminal-to-criminal supply chains that enable attacks.
Some of these were previously studied from the direct attackers and victims' perspectives, such as romance scams~\cite{10.1007/978-3-319-20550-2_12}. Our new understanding of the underlying supply chains can illuminate different and potentially more effective methods of undermining these threats~\cite{Levchenko:2011}.


\begin{figure*}[!ht]
 \centering
 \includegraphics[scale=0.7]{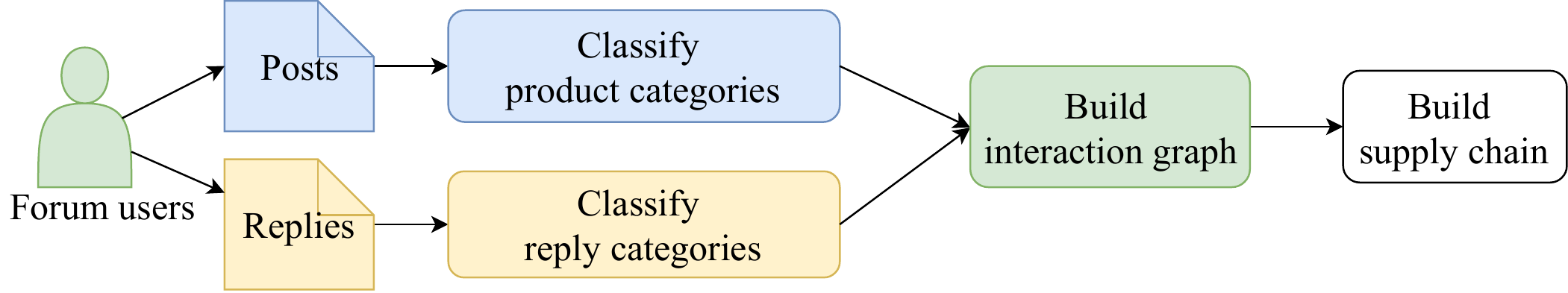}
 \caption{Our supply chain detection approach.}
 \label{fig:approach}
\end{figure*}

\begin{figure*}[ht]

 \centering
 \includegraphics[scale=0.7]{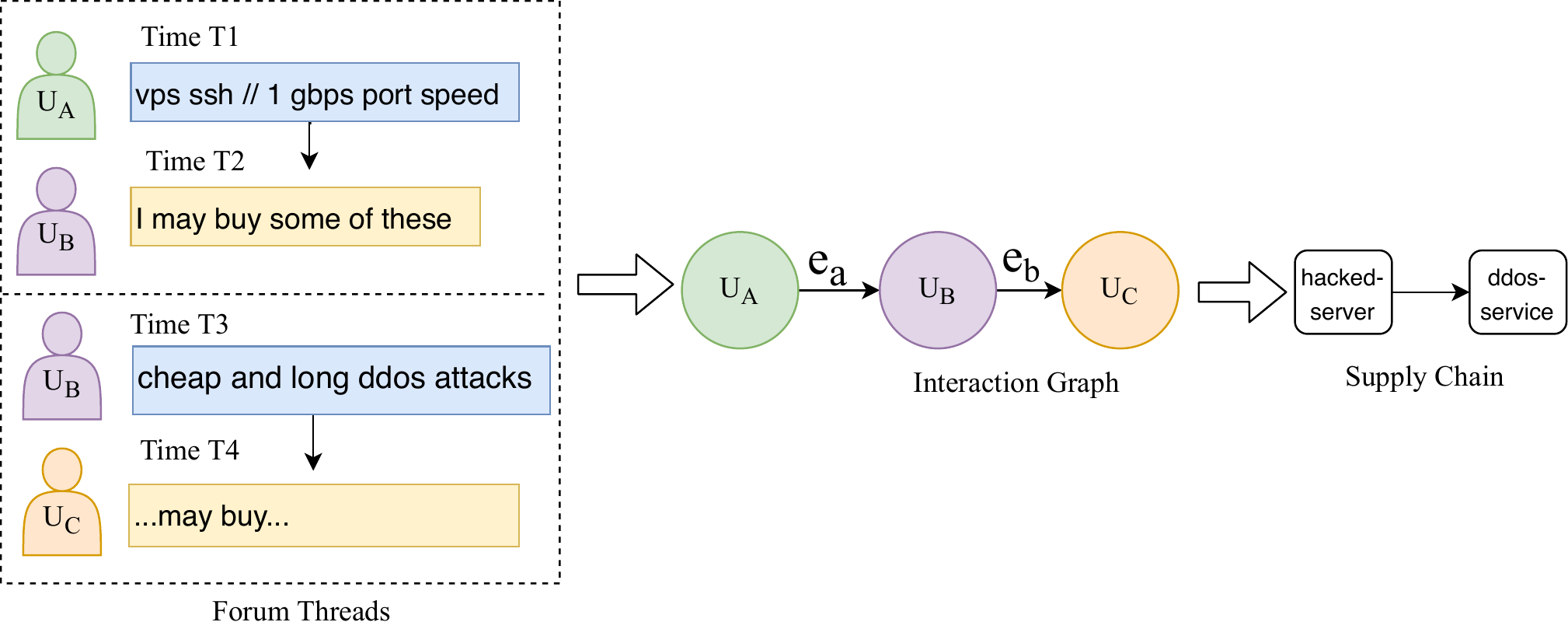}
 \caption{Example of constructing supply chains from two \hf\ threads.}
 \label{fig:scl}
\end{figure*}

\section{Forums}
\label{data}

To evaluate our approach, we chose two popular forums: Antichat (Russian) and Hack Forums (English) (Table~\ref{table:descriptive_statistics}).
We chose these forums because they are large, publicly available, and have been used to evaluate cybercrime analysis methodologies in prior studies~\cite{Durrett2017IdentifyingPI,portnoff2017tools}. We also have access to more recent data from these two forums and other large cybercrime forums through an agreement with a security company. However, this data is private and our agreement prevents us from publicly sharing these datasets. Thus, in this paper we chose to limit our evaluation to public datasets so as to enable reproducibility of our major findings.


\paragraph{\hf}
Hack Forums is a major English-language forum covering many cybercrime-related topics. The forum has been active since 2007. We use a partial 6 year scrape between April 2009 and April 2015. The scrape is partial because it only includes sell posts from the Hack Forums Marketplace. 

\paragraph{Antichat}
Antichat is a major Russian-language forum covering cybercrime-related topics. Examples include password cracking, stolen accounts, and physical weaponry. We obtain a full database leak containing posts between January 2005 and June 2010.


\subsection{Availability and Ethics}
The two data sources used in this study are a scrape of public postings from \hf\ and publicly leaked dump of Antichat cybercrime forums. Both of these datasets are publicly available data and have been analyzed in prior studies~\cite{Durrett2017IdentifyingPI,portnoff2017tools}. The validity of this data has been established by these prior studies.
Our analysis focuses on creating and evaluating generalizable methods for constructing supply chains. We did not attempt to analyze any Personally Identifiable Information (PII) in these datasets. All case studies mentioned in section \ref{case_studies} are anonymized except for those that have been previously publicly reported, and so are any analysts mentioned. This study was exempted by our Institutional Review Board (IRB) since the datasets were publicly available and the focus of our study was not on analyzing PII in the dataset. Our methods conform to recommended best practices for ethical research pertaining to datasets of illicit origin~\cite{Thomas:2017:EIR:3131365.3131389}.

In order to facilitate reproducibility, we will make available upon request to other researchers all of the data, annotations and code for our study. However, we will not make the data publicly available in a way where it might be indexed by search engines and increase the possible harm from any PII contained in these datasets.


\section{Approach}
\label{approach}

Our goal is to automatically build a supply chain for a criminal forum by analyzing the posts and replies. To build a supply chain, we need a chronological record of which  products the users of a forum are buying and selling. We use automated classifiers to categorize the posts and replies, then build an interaction graph and use the graph to build the supply chain (Figure~\ref{fig:approach}).

\subsection{Classify Product}
To find a supply chain, we need to first identify the categories of the product bought or sold in forum posts. We use supervised classification to classify products into $n$ categories, where the specific categories might depend on the forum and the analyst. The details of our classifier will be presented in section~\ref{eval}.

\subsection{Classify Replies}
Once we have the product categories, we need to identify who bought or sold the products to build a supply chain. The reply classifier is similar to the product classifier. It is a supervised classifier that uses TF-IDF of character n-grams to classify replies into three main categories: \texttt{buy}, \texttt{sell} and \texttt{other}. To classify only relevant replies, we run a quote removal algorithm to remove instances of a reply quoting a previous reply before feeding it to the reply classifier.




\subsection{Build an Interaction Graph}
To determine who bought what and when, we build an interaction graph for a forum. The interaction graph is a directed graph, $G = (U, E)$ where each node $u \in U$ is a user who posts on the forum, and each edge $(u_a, u_b) \in E$ indicates that user $u_a$ sold a product to user $u_b$. 

To build this graph, we use the product and reply classifiers. For a post by $u_a$, the product classifier determines the category of the product sold in the post and the reply classifier determines whether user $u_b$'s  reply to $u_a$'s post implies buying the product. We also consider the time of the buy reply as the time at which the user $u_b$ purchased the item from user $u_a$.

\subsection{Build Supply Chain}
The purpose of a supply chain graph is to illuminate the sequence of processes involved in various criminal activities. 
The supply chain graph of a forum is a directed graph, $S = (C, I)$, where 
each node, $c_i \in C$, is a product category and 
each edge, $(c_a, c_b) \in I$, indicates that at least one user in the forum bought a category  $c_a$ product and sold a category  $c_b$ product.
We use  breadth-first search on the interaction graph to create the supply chain.  Figure~\ref{fig:scl} shows an example of creating a supply chain from two Hack Forums threads. 

To create the supply chain graph, we define a supply chain link in an interaction graph as a tuple of two interactions $e_a$ and $e_b$, where $e_a$ is an edge from user $u_{a}$ to user $u_{b}$, and $e_b$ is an edge from $u_{b}$ to $u_{c}$. This means that user $u_{a}$ sold a product to user $u_{b}$, who then sold a product to user $u_{c}$. Our breadth-first search follows the supply chain links in the interaction graph in chronological order (Algorithm~\ref{alg}).




%
When adding links to the supply chain graph, we do not want users who are outliers disproportionately buying or selling certain items to unfairly add to their links. For example, somebody might buy 100 items and then sell once, adding 100 links (an exaggeration of the problem). We mitigate this issue by dividing the weight that each user contributes to each link by the total number of links to which that user contributes. Our method of attenuation is: If a user appears in $n$ edges, then that user adds $ 1 / n $ to each edge in which they appear; so, each user adds a total of $1$ to the entire graph, and each link still appears, but contributes less than if that user only created one or a few links with a single purchase. These weights after attenuation are used later in section~\ref{analysis}. For example, if user $A$ sold one product to user $B$ and then user $B$ sold 50 products to other buyers this would be attenuated to only a single supply chain. 


\begin{algorithm}[!ht]
\DontPrintSemicolon
\SetAlgorithmName{Supply Chain Algorithm}{}
\textbf{Input:} Interaction graph, $G = (U, E)$ where $u \in U$ $\leftarrow$ user and $(u_a, u_b) \in E$ $\leftarrow$ user $u_a$ sold to user $u_b$

\textbf{Output:}  Supply chain graph, $S = (C, I)$, where 
$c_i \in C \leftarrow$ product category and
$(c_a, c_b) \in I \leftarrow$ users bought a $c_a$ product and sold a $c_b$ product


  \While{not every user $u \in U$ has been discovered}{
         $L1 \leftarrow$ undiscovered user $u \in U$ \;
         \While{$L1$ is not empty}{
            $L2 \leftarrow$ empty list 
            
            \For{each user $u_i \in L1$}{
                \For{each undiscovered user $u_j$ who sold to $u_i$}{
                    $W \Leftarrow$ number of items $u_j$ bought and then sold \;
                    \For{each undiscovered user $u_k$ who sold to $u_j$}{
                        
                        \If{$W > 0$}{
                            $(c_a, c_b)\Leftarrow$ supply chain link between ($u_i$, $u_j$) and ($u_j$, $u_k$), divided by $W$  \;
               $I \Leftarrow$ $(c_a, c_b)$ $\bigcup$ $I$ \;
                            $L2 \Leftarrow$ $u_j$ $\bigcup$ $L2$ \;
                        }
                    }
                }
            }
            $L1 \Leftarrow L2$ \;
         }
    }
 \caption{Modified Breadth-First Search for Supply Chain Generation}
 \label{alg}
 
\end{algorithm}

\section{Evaluation}
\label{eval}

To evaluate our approach, we first label each post from our Antichat and \hf\ datasets into product categories and each reply into reply categories. Using our labeled data as groundtruth, we evaluate the performance of the classifiers and end-to-end supply chain link algorithm. 

\subsection{Labeling Ground Truth}
We perform two types of labeling: Product labeling and Reply labeling. Unlike prior work that annotates the head of a noun phrase describing a product~\cite{Durrett2017IdentifyingPI,portnoff2017tools}, we label each post into one of the predetermined categories. This approach is less time consuming. Moreover, a finer grained product identification might be counterproductive for the analysis of supply chains, as a large number of categories will result in complicated visualizations.  

\paragraph{Product Labeling}
We identified 14 product categories for our datasets (Table \ref{table:product_annotation_labels}). These categories were determined by domain experts based on reading posts in both forums and choosing products of interest in line with their analysis goals. 
To adapt our classifiers to other forums, an analyst can modify the categories to fit the forum.
An alternative way of choosing product categories might be to explore unsupervised clustering methods~\cite{pmlr-v28-kim13,Zhang:2018}.

The product categories were annotated by domain experts who have native fluency of the forum's primary language. The distributions of the annotations per source are shown in Table \ref{table:product_annotation_labels}. Figure \ref{fig:other_posts} in the appendix shows examples of what a post classified \texttt{other} looks like. The \texttt{other} posts could not fit into any of the product categories we already identified. 

Both annotated datasets are highly unbalanced making the classification task particularly hard. Unlike for \hf, however, our initial random Antichat data-sample had more \texttt{account} posts than \texttt{other}. The ramification of this is that when the classifier struggles to decide how to classify data points it tends to put it into the \texttt{account} class. Since we decided to prioritize precision for all of the classes, except \texttt{other}, we decided to undersample \texttt{account} to be smaller than \texttt{other} in our training sets. The effect of this is an increase in precision of the \texttt{account} class but a decrease in recall. The numbers we report for Antichat in Table~\ref{table:product_annotation_labels} are after undersampling the \texttt{account} class so that it is smaller than the \texttt{other} class. This undersampled data was only used for training our models. The natural distribution was used for constructing testing sets and analysis.

\begin{table*}[htb!]
\centering
\begin{tabular}{lp{4in}cc}
\toprule
\textbf{Product}   & \textbf{Description} & \textbf{Antichat} & \textbf{\hf} \\ \midrule
\vspace{0.07cm}
Account            & Selling or requesting an account, multiple accounts, or access codes. This also includes account creation automation software.                                  & 19\%              & 19\%              \\
\vspace{0.07cm}
Botnet             & Selling or renting access to computers infected with malicious software.                                                                                          & 2\%               & 1\%                 \\
\vspace{0.07cm}
Crypter            & A piece of software which obfuscates malware.                                                                                                                      & 2\%               & 6\%                 \\
\vspace{0.07cm}
DDoS service       & Selling or requesting a DDoS attack.                                                                                                                                & 1\%               & 4\%                 \\
\vspace{0.07cm}
Hacked server      & Selling or requesting a single hacked server.                                                                                                                        & 18\%              & 1\%                 \\
\vspace{0.07cm}
Hack-for-hire      & Offering targeted hacking, malware coding or requesting a specific service.                                                                                            & 4\%               & 7\%              \\
\vspace{0.07cm}
Hosting            & Hosting a website, game server, or otherwise maintaining it.This includes DDoS mitigation.                                                                              & 3\%               & 4\%                 \\
\vspace{0.07cm}
Malware            & A piece of malicious software that is executed on a victim's machine. Examples of this include cryptocurrency miners and ransomware.                                 & 8\%               & 8\%               \\
\vspace{0.07cm}
Proxy              & Selling or requesting a proxy/VPN.                                                                                                                                    & 3\%               & 1\%                 \\
\vspace{0.07cm}
Social booster     & Supports gaining social media attention. Examples of this are, ``buying likes/views'', ``selling a twitter followers''.                                             & 3\%               & 3\%                 \\
\vspace{0.07cm}
Spam tool          & Selling or requesting an email/chat service spam tool or spamming service.                                                                                         & 9\%               & 1\%                 \\
\vspace{0.07cm}
Traffic            & Selling real or fake visitors to a site. Does not include social media related ``traffic''.                                                                     & 5\%               & 1\%                 \\
\vspace{0.07cm}
Video game service & Selling or requesting any service related to video games. Includes things like mods, points, and power-leveling. Does not include selling video game accounts. & 1\%               & 10\%               \\
\vspace{0.07cm}
Other              & Anything that doesn't fall into the previous categories.                                                                                                       & 22\%              & 34\%               \\ \bottomrule
TOTAL              &                                                                                                                                                                & 18,796            & 14,557              \\ \bottomrule
\end{tabular}
\caption{Product annotation labels and distribution per source}
\label{table:product_annotation_labels}
\end{table*}

\paragraph{Reply Labeling}

We label  each reply into three categories: buy, sell and other. The distribution of reply types is highly dependant on the structure and rules of the forum as shown in Table~\ref{table:reply_annotation_labels}. For example, there are not many \texttt{sell} replies on Hack Forums relative to the other categories. Replies labelled as \texttt{other} tend to be questions about products and informational.

\begin{table*}[htb!]
\centering
\begin{tabular}{@{}llcc@{}}

\toprule
\textbf{Reply Type} & \multicolumn{1}{c}{\textbf{Description}}                                                                                                                                                                           & \textbf{Antichat} & \textbf{Hackforums} \\ \midrule
Buy                                     & Someone wants to buy or bought a product.                                                                                                                                                                       & 8\%               & 12\%                 \\
Sell                                 & Someone making a sale offer to the original poster of a thread.                                                                                                                                                    & 8\%               & 2\%                 \\
Other                                   & Anything that didn't fall into the previous categories.                                                                                                                                                            & 84\%             & 86\%               \\ \bottomrule
TOTAL                                   &                                                                                                                                                                                                                    & 9,992             & 5,898               \\ \bottomrule
\end{tabular}
\caption{Reply classification labels and distribution per source}
\label{table:reply_annotation_labels}
\end{table*}

\begin{table*}[ht]
\centering
\begin{tabular}{@{}lrllcllclllll@{}}
\toprule
                             & \multicolumn{6}{c}{\textbf{Antichat}}                                                                                                                 & \multicolumn{6}{c}{\textbf{\hf}}                                                                                                               \\ \midrule
                             & \multicolumn{3}{c}{\textit{Product}}                                           & \multicolumn{3}{c}{\textit{Reply}}                                   & \multicolumn{3}{c}{\textit{Product}}                                 & \multicolumn{3}{c}{\textit{Reply}}                                             \\
\textbf{Model}               & \multicolumn{1}{c}{Prec} & \multicolumn{1}{c}{Recall} & \multicolumn{1}{c}{F1} & Prec           & \multicolumn{1}{c}{Recall} & \multicolumn{1}{c}{F1} & Prec           & \multicolumn{1}{c}{Recall} & \multicolumn{1}{c}{F1} & \multicolumn{1}{c}{Prec} & \multicolumn{1}{c}{Recall} & \multicolumn{1}{c}{F1} \\ \midrule
\textbf{FastText}           & 0.824                    & 0.734                       & 0.764                  & 0.823          & 0.318                       & 0.450                  & 0.722          & 0.582                       & 0.627                  & 0.800                   & 0.427                        & 0.539                  \\
\textbf{Logistic Regression} & 0.831                    & 0.718                       & 0.753                  & \textbf{0.874}          & 0.245                       & 0.381                  & 0.617          & 0.564                       & 0.617                  & \textbf{0.852}                    & 0.363                       & 0.492                  \\
\textbf{SVM}                 & 0.817                    & 0.748                       & 0.767                  & 0.654          & 0.213                       & 0.301                  & 0.716          & 0.578                       & 0.614                  & 0.812                    & 0.332                       & 0.440                  \\
\textbf{XGBoost}             & \textbf{0.824}                    & 0.677                       & 0.729                  & 0.713          & 0.227                       & 0.328                  & \textbf{0.734}          & 0.577                       & 0.627                  & 0.819                    & 0.352                       & 0.465                  \\ \bottomrule
\end{tabular}
\caption{Weighted precision, recall and F1 scores of classifiers across datasets and tasks, with stratified k-fold cross-validation}
\label{table:precision_f1_scores_strat}
\end{table*}

\begin{table}[h]
\centering
\begin{tabular}{@{}lrr@{}}
\toprule
                     & \multicolumn{1}{c}{\textbf{Antichat}} & \multicolumn{1}{c}{\textbf{\hf}} \\ \midrule
\textbf{Top Reply}   & sell (0.890)                       & sell (0.893)                         \\
\textbf{Low Reply}   & buy (0.860)                        & buy (0.828)                          \\
\textbf{Top Product} & crypter (0.905)                       & crypter (0.832)                         \\
\textbf{Low Product} & other (0.502)                         & spam-tool (0.497)                       \\ \bottomrule
\end{tabular}
\caption{Highest and lowest categories across classifiers in terms of precision}
\label{table:precision_performance}
\end{table}

\subsection{Validating Product Category and Reply Classifier}

We extract features from posts using TF-IDF~\cite{Baeza-Yates:1999:MIR:553876} to produce a vector where each element corresponds to that term's TF-IDF score.
We experiment with character and word n-grams. We chose character n-grams as it outperformed word n-grams on our datasets. 
We tested four classifiers: 1) FastText, 2) Logistic Regression, 3) SVM, and 4) XGBoost. We tuned each of these methods to reduce over-fitting of the labeled data~\cite{Chen:2016:XST:2939672.2939785,DBLP:journals/corr/JoulinGBM16}. We selected these four classifiers as a diverse set of classifier types but we did not test an exhaustive set of classification methods. 

For our product category classifier evaluation, we used the labeled data described in Table~\ref{table:product_annotation_labels} and classified each post into one of 14 categories. We perform stratified 5-fold validation of each classification algorithm since our classes are highly imbalanced. For highly imbalanced datasets, regular k-fold cross validation often produces a biased evaluation because the limited number of folds generated can have a class distribution that does not match the one in the actual data. We use stratified k-fold validation since it ensures an ``apples-to-apples'' evaluation where the same distribution of the target values that exist in the main data set are maintained for each fold~\cite{SCV}.



In order to select the classifier used in our analysis, we use a weighted average of the precision scores across all the categories except \textit{other}. We ignore the \textit{other} class in this metric because we are not concerned with having a low precision in that category, since posts classified as \textit{other} will be filtered out by our supply chain identification algorithms; our goal is to optimize the percentage of posts used in the supply chain algorithm that truly belong in their given category. Although other classifiers perform equally well as XGBoost in weighted F1 scoring, we choose XGBoost for all product classification tasks because it performs better in our weighted non-\textit{other} precision scoring metric providing a precision of 0.824 on Antichat and 0.734 on \hf. We see the performance when considering weighted precision in Table \ref{table:precision_f1_scores_strat}. The remaining Antichat and \hf\ product posts were classified using the XGBoost models.

The second classification task used for identifying supply chains is to categorize each reply to the first post in each thread into one of three categories: \textit{buy}, \textit{sell}, or \textit{other}. We chose the categories based on the needs of the supply chain algorithm. For our reply category classifier evaluation, we used the labeled data described in Table~\ref{table:reply_annotation_labels}. We again perform stratified 5-fold validation of each classification algorithm since our classes are highly imbalanced.

Similarly a single classifier outperforms the rest in reply classification, as seen in Table \ref{table:precision_f1_scores_strat}. By our weighted non-\textit{other} precision metric, Logistic Regression performed the best across both datasets, providing 0.874 precision on Antichat and 0.852 precision on \hf. Although we care about overall model performance, we want to ensure the integrity of our supply chain links, described later, and a higher precision would ensure that. In other words, we decided to trade recalling potential classification for the legitimacy of the classifications the models made.

Lastly, Table~\ref{table:precision_performance} depicts which classes across classifications tasks performed the best and worst in terms of precision. We theorize that \texttt{sell} performs the best in reply classification because the wording of that text is more distinct than \texttt{other} and \texttt{buy}. We argue a similar theory is likely for the \texttt{crypter} product in both product classification tasks because the posts advertising these products often use many of the same words.

\paragraph*{Limitations in Classification}
We tested four classifiers with a small set of features over two forums. Our classifiers could potentially be improved by identifying more useful features, exploring additional classifiers, and diversifying the forums used for testing. Our product category taxonomies are also specific to the two forums that we analyzed and would need to be modified for other forums. 

\label{supply}

\subsection{Validating Supply Chain Link }
We validate whether the chain link tuples output by our algorithm describe ``true'' links in supply chains. A ``true'' link is what we would consider a link in a supply chain, not simply a coincidence where a user purchased something and then sold something else, unrelated to the item they purchased, or an error in classification resulting in the lack of a purchase or even a product. For example, a user purchasing a program that adds followers to any Instagram account, and then subsequently selling a popular Instagram account is a ``true'' link. To evaluate this, we manually check the links produced by the algorithm.

\begin{table*}[!t]
\centering
\begin{tabular}{@{}llccllcc@{}}

\toprule
\textbf{Link Type} & \multicolumn{2}{c}{\textbf{\hf}} &  \multicolumn{2}{c}{\textbf{Antichat}}  \\ \midrule
Link Type & Algorithm Output & Sample Baseline & Algorithm Output & Sample Baseline \\ \midrule
Related & 37 (31\%) & 4 (10\%) & 181 (41\%) & 18 (22\%) \\
Resell & 6 (5\%) & 1 (3\%) & 77 (17\%) & 12 (14\%) \\
Unrelated & 34 (28\%) & 1 (1\%) & 82 (18\%) & 26 (31\%) \\
Lack of product & 15 (12\%) & 0 (0\%) & 57 (13\%) & 23 (28\%) \\
Lack of purchase & 28 (24\%) & 30 (86\%) & 43 (10\%) & 4 (5\%) \\
\bottomrule
TOTAL & 119 & 35 & 441 & 83 \\ \bottomrule
\end{tabular}
\caption{Attenuated Link Truth Level by Forum. ``Algorithm Output'' provides attenuated values for links detected by our complete method. ``Sample Baseline'' gives the same for links detected by only the Supply Chain Algorithm~\ref{alg} without using the results of the reply classifier to filter links (i.e., without limiting links to those created by ``buy'' replies). \texttt{Related} links have a user who purchased a product and then sold another product likely using the previous one in a supply chain fashion; (e.g., buying a hacked server and then selling DDoS attacks). \texttt{Resell} links have a user who purchased and resold the same product, possibly with an updated description. \texttt{Unrelated} links have a user who purchased a product and sold another unrelated product that is not likely to logically result from the source product. \texttt{Lack of product} happens when the product classifier determines a post belongs in a product category, and feeds it to the Supply Chain Algorithm, but the post was not actually discussing selling a product of interest (i.e., the product classifier labeled it as a product when in truth it is ``other''). \texttt{Lack of purchase} happens when the linking user did not actually indicate purchasing the source product, so the link detected cannot be a supply chain link (i.e., the reply classifier labelled it as a ``buy'' reply, but in truth it was not). Attenuation is performed to counter the effect of a user either buying or selling multiple times, which would otherwise result in many links when combined.}
\label{table:link_relevance}
\end{table*}

Table~\ref{table:link_relevance} shows the attenuated counts and percentages of links that were an example of a forum member buying a product and then selling another product that is produced using the previous product in a supply chain fashion (\texttt{related}), someone buying and then reselling the same item with a slightly changed description (\texttt{resell}), and a user purchasing one product and selling another where the two products are not related in any reasonable supply chain scenario (\texttt{unrelated}). Table~\ref{table:link_relevance} also displays attenuated counts and percentages of links in which the product classifier determined that a post belonged in one of our product categories; but in truth, that post was not actually discussing selling a legitimate product (\texttt{Lack of product}), or links with errors in reply classification made it so that the link detected cannot be a supply chain link because the user who replied did not purchase the product (\texttt{Lack of purchase}). Out of these, links classified as \texttt{related} and \texttt{resell} are considered proper supply chain links. Our algorithm outputs 31\% related and 5\% resell in \hf\ and 41\% related and 17\% resell in Antichat.

``Algorithm Output'' shows the attenuated counts and associated percentages of links detected by our complete method that fell into each Relevance Level. ``Sample Baseline'' shows the same detected by only the supply chain algorithm (the modified breadth-first search), without using the results of the reply classifiers to filter links (i.e., without limiting links to those created by ``buy'' replies). 

Attenuation is performed to counter the effect of a few outliers either buying or selling multiple times, which would otherwise result in many links. The amount that each link contributes to its respective relevance class is divided by the total number of links that the linking user creates. For example, if a user purchases one item and then sells many items, resulting in many links, we want that user to contribute less to each link validation type (mentioned in Table~\ref{table:link_relevance}) where their links belong than if they only created one or a few links with the single purchase. If we don't use attenuation, then outlier users will affect the distribution of link validation types.

To determine if using our classifiers (to filter out \texttt{other} product posts and limit to \texttt{buy} replies) increases the density of valid links by disproportionately filtering out invalid links, we manually validated all of the links we discovered after filtering. For a baseline comparison, we used the supply chain algorithm to find links, but without filtering using the results of the reply classifier. We then manually evaluated a random sample of 100 of these unfiltered supply chains (note the number of links after attenuation is 35 for \hf\ and 83 for Antichat). The results of our supply chain evaluation are in Table \ref{table:link_relevance} as ``Sample Baseline.'' Considering ``related'' and ``resell'' links as relevant, we find that our classifiers improves the rate of relevant links from 13\% to 36\% for \hf\ and from 36\% to 58\% for Antichat. It is infeasible to compute the recall since there is no efficient method for identifying all relevant links in our datasets. 

A statistic of note in Table~\ref{table:link_relevance} is that 86\% of \hf\ sample baseline links are ``Lack of purchase''; the vast majority of these links were made using ``vouching'' replies. A vouching reply is one in which the author vouches for the original poster, claiming that they have worked with that user and that the original poster can be trusted. This is likely because the \hf\ community places importance on vouches to determine a user's trustworthiness, so gangs of users post vouching replies to each others' posts, creating links.

\paragraph*{Limitations in Supply Chain Generation}
The main limitation evident from this evaluation is that not every tuple is a legitimate link, requiring manual exploration of the links detected. However, we show that our algorithm finds fewer non-legitimate links with the classifiers than without. This issue, however, is indicative of an overarching limitation, which is that we cannot be sure that all replies marked as ``buy" actually resulted in a transaction. For example, a user bidding on an item was marked as a user wanting to purchase, but it is possible someone outbid them, and they did not actually receive the product. 

Other limitations result from the timing part of our definition of supply chains. For example, sometimes limiting supply chain links to those where the user purchased the source product before selling the destination product is an incorrect constraint. For instance, a user could wait to buy a required product for their offering until they have a customer. It may also increase the percentage of relevant supply chain links to have an upper limit for the amount of time between interactions in a tuple; however, we then miss the cases when there is a large delay.

Another limitation is that our notion of related products is limited by our ability to correctly interpret these forum messages and infer connections. Thus, we might have missed some new supply chains because we did not deduce the connection. However, we place an emphasis on precision to reduce false positives at the cost of additional false negatives.
\begin{figure*}[!ht]
 \centering
 \includegraphics[width=\textwidth]{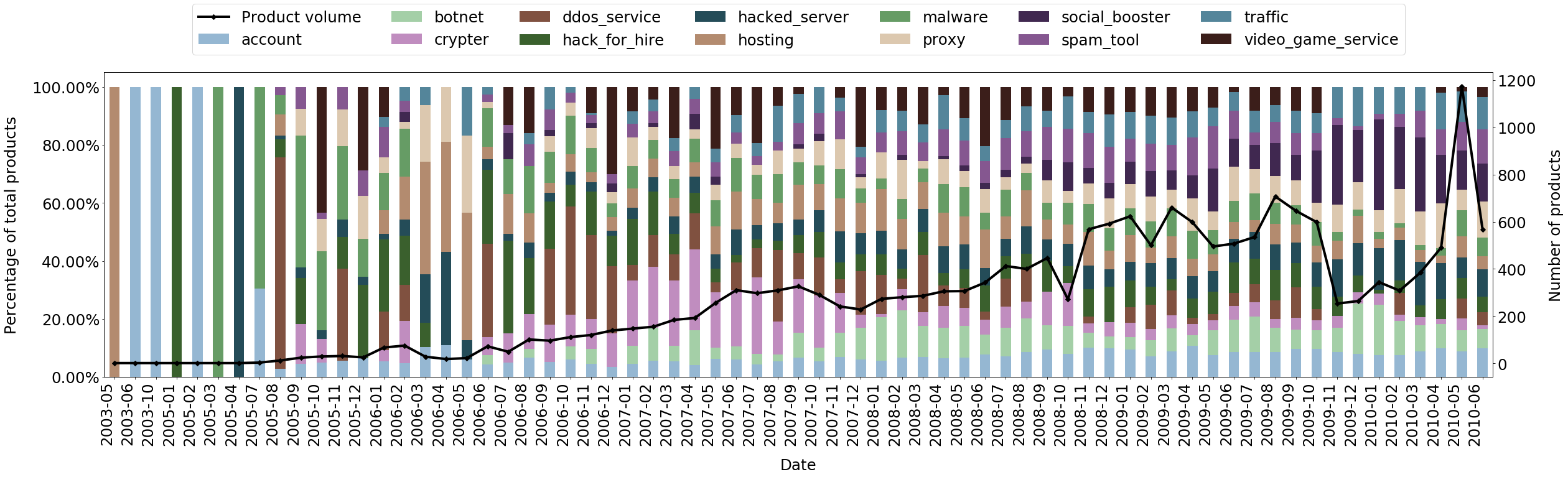}
 \caption{Antichat products trends. 73,115 posts in total. The right hand y-axis is the actual volume of posts, and the left y-axis is the percentage of the total number of posts. The black line signifies the volume of post and is associated with the right y-axis. In Antichat, \texttt{accounts} and \texttt{hosting} were most popular until the ecosystem diversified into mixed monthly degrees of our 14 product taxonomy.}
 \label{fig:antichat_products}
\end{figure*}

\begin{figure*}[!ht]
 \centering
 \includegraphics[width=\textwidth]{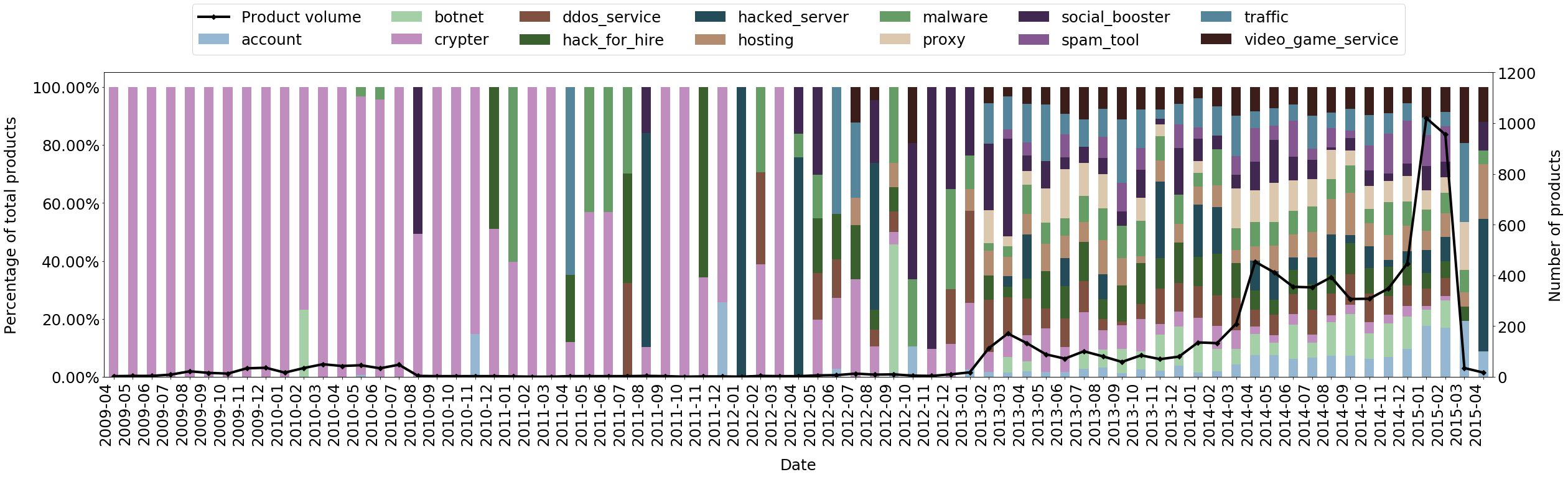}
 \caption{\hf\ products trends. 14,447 posts in total. The right hand y-axis is the actual volume of posts, and the left y-axis is the percentage of the total number of posts. The black line signifies the volume of post and is associated with the right y-axis. In \hf, product trends demonstrate a \texttt{crypter} beginning signifying the need to obfuscate malware/software. This need became less dominant as the ecosystem become more diverse.}
 \label{fig:hackforums_products}
\end{figure*}

\section{Analysis}
\label{analysis}

This section contains analyses focused on providing useful information to an analyst investigating underground forums. We show how insight from these analyses can make an analyst more efficient at detecting shifts in threats and understanding general ecosystem trends. All the analysis in this section is based on data classified by the best performing model for each task: XGBoost for product classification and Logistic Regression for reply classification. In every classification task, aside from products in \hf, the model was trained on all the annotated posts and the remaining posts were predicted. 

The \hf\ product task is unique because we annotated all the posts. In order to reclassify them, we performed k-fold classification where all of the annotated product posts were split into 5 folds. We then classified all of the posts in one fold using a model trained on posts from the remaining four folds and repeated this until all of the posts in the five folds were classified and an XGBoost model trained on four folds was used to predict the remaining fold. This was repeated until each fold was predicted. 
We classified the \hf\ posts using this method so that we could provide a realistic end-to-end assessment of our supply chain detection method that included the likely misclassification error, assuming it is not possible to label all of the posts.

\subsection{Product Analysis}

We performed time series and statistical analyses of the classified and annotated product posts using the taxonomy from Figures~\ref{fig:antichat_products} and \ref{fig:hackforums_products}. The dataset analyzed is the one described in Table~\ref{table:descriptive_statistics}, with 73,115 threads in Antichat over 7 years, and 14,447 threads in \hf\ over 6 years. Of these, all of the 14,447 \hf\ posts and 21,996 of the Antichat posts were manually annotated; the rest were classified using an XGBoost classifier trained on the annotated posts. We demonstrate how a product-level trend analysis gives insight into what activity is present on a forum and how forums change over time. Normally this requires manually reading through hundreds of posts to get a sense of changes.

The first difference is between the initial illicit activity of the two forums. Our Antichat data begins in Summer 2008 where the initial product offerings focused either on malicious hosting and accounts.


\hf, on the other hand, shows very different beginnings, where most of the data we classified pertains to \texttt{crypters}. This means most individuals were buying/selling items related to obfuscating software. This implies a very different story than Antichat, where there was likely initially a demand for this single product. As time progresses, we see \hf\ become a more diverse marketplace, where \texttt{crypter} related products become a fairly insignificant part of new offerings. 

The interesting similarity between the two forums is that as volume increases we see that the product offerings become more diverse. This indicates that these forums evolve into ecosystems where specialized and likely more efficient sellers start to organize into supply chains where one seller of a higher level service, such as DDoS attacks, depends on hacked servers supplied by other sellers. Similar to normal business ecosystems, this likely enables increasingly efficient and sophisticated attacks to emerge. These product category trend analyses only show what is being sold but they do not illuminate the connections between products.

\begin{figure}[!ht]
\begin{center}
\includegraphics[width=0.45\textwidth]{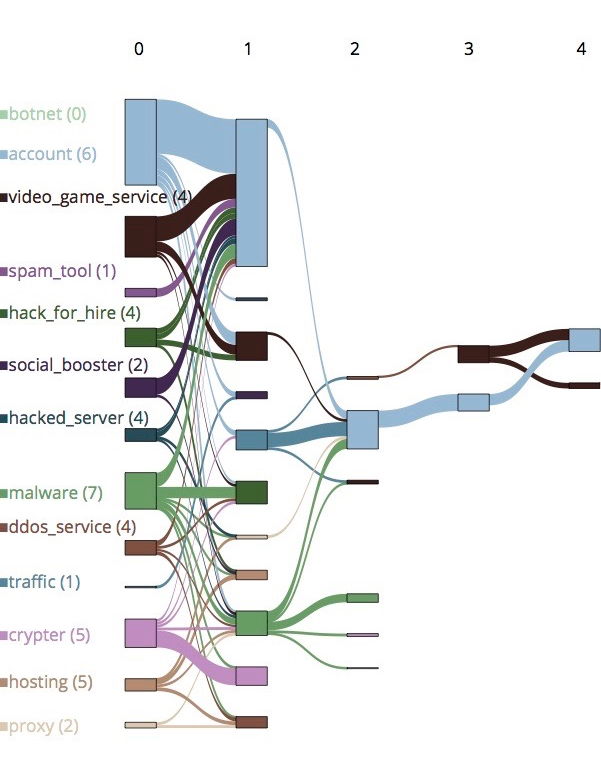}

\caption{Supply chains found in \hf\, limited to links manually validated as ``Related'' or ``Resell''. Edges are colored according to source product category, and have widths determined by the number of users who purchased the source product and sold the destination product. Numbers at the top correspond to the level in the modified breadth-first search algorithm at which the node was discovered. The number of chains originating with each product category is denoted next to the product category names. Each user contributes to each link the amount determined by our method of attenuation. We omitted for space two chains which were longer than level 4.}
\label{fig:Hackforums_Relevant_Alluvial}
\end{center}
\vspace{-0.5cm}
\end{figure}

\begin{figure*}[!ht]
\begin{center}
\includegraphics[width=0.95\textwidth]{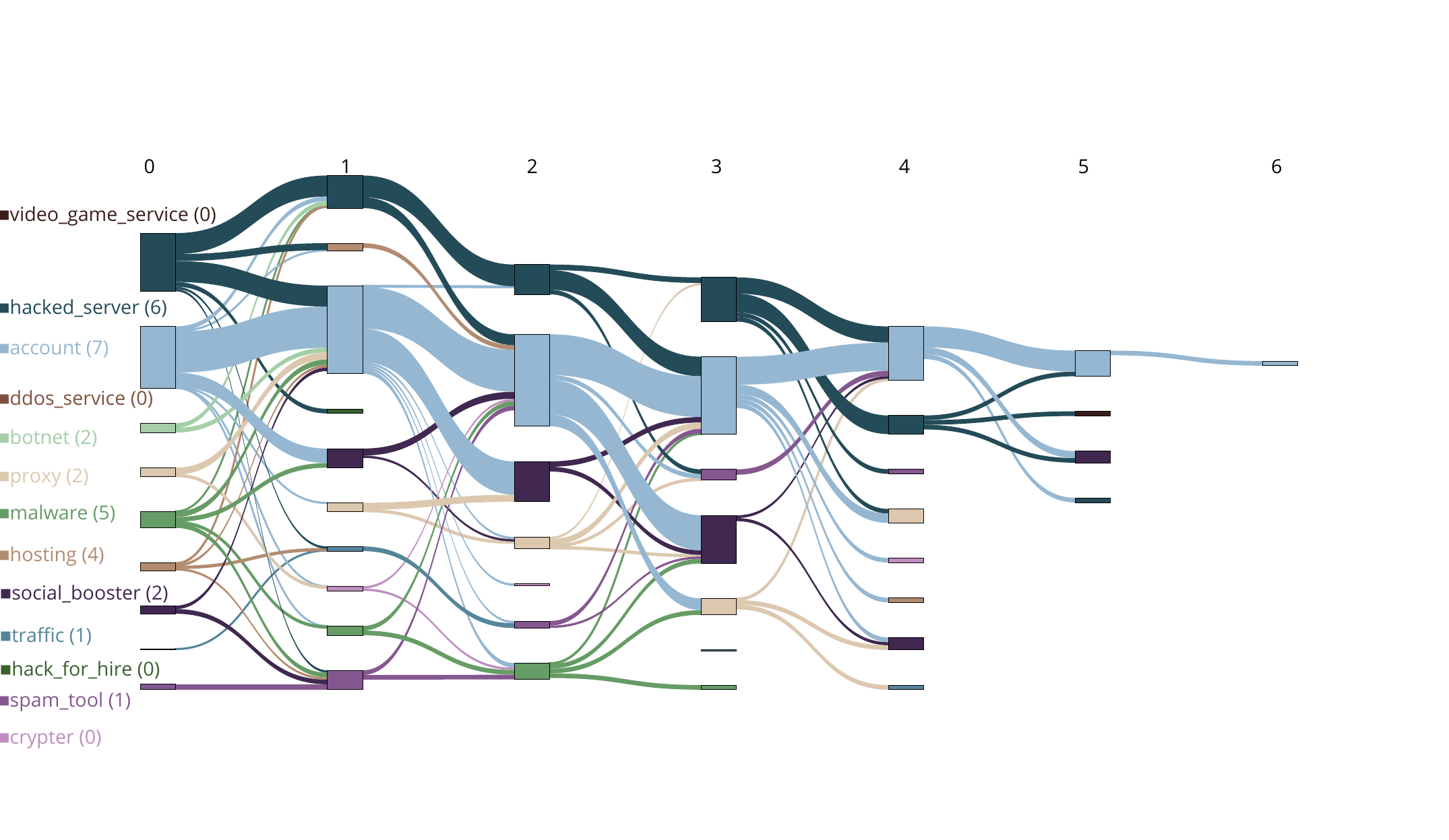}
\vspace{-0.5cm}
\caption{Supply chains found in Antichat, limited to links manually validated as ``Related'' or ``Resell''. Edges are colored according to source product category, and have widths determined by the number of users who purchased the source product and sold the destination product. Numbers at the top correspond to the level in the modified breadth-first search algorithm at which the node was discovered. Number of chains originating with each product category are denoted next to the product category names. Each user contributes to each link the amount determined by our method of attenuation.}
\label{fig:Antichat_Relevant_Alluvial}
\end{center}
\vspace{-0.5cm}
\end{figure*}

\subsection{Supply Chain Analysis}

To give a sense of the scale, the total number of posts used was 14,557 and 51,119 for \hf\ and Antichat, respectively. Out of these, only posts with products classified into categories outside \texttt{other} were used for supply chain analysis; this was 9,601 for \hf\ and 28,991 for Antichat. The total number of links, when allowing posts with \texttt{other} products is 1,041 and 9,043 links in \hf\ and Antichat, respectively, and when filtering out \texttt{other} product posts is 233 and 1179 in \hf\ and Antichat, respectively. These same links with filtering out \texttt{other} products, through attenuation, become 119 for \hf\ and 441 for Antichat.


We produced alluvial graphs of the supply chains resulting from our algorithms only  including links manually labeled as \texttt{related} or \texttt{resell}. Figures~\ref{fig:Hackforums_Relevant_Alluvial} and \ref{fig:Antichat_Relevant_Alluvial} depict these supply chains as alluvial graphs for \hf\ and Antichat, respectively. There are 119 links in \hf\ and 441 links in Antichat, after attenuation.


In both \hf\ and Antichat, \texttt{accounts} as products are central. In Antichat, the product that feeds most into \texttt{accounts} is \texttt{hacked-server}, followed by \texttt{malware} and \texttt{spam-tool}. The product most often produced by members who purchased \texttt{accounts} is \texttt{social-booster}. This makes sense according to domain experts, since hacked servers are used to brute-force and create accounts, and acquiring a large number of accounts helps to generate spam, social media likes, clicks, etc. In \hf, \texttt{video-game-service} and \texttt{accounts} are heavily linked due to many of \hf's users being avid gamers. Links from both forums are discussed in more detail as case studies.

\section{Case Studies}
\label{case_studies}

In this section, we demonstrate how an analyst can use the supply chains to explore the criminal markets, understand how products are derived, and what popular supply chains say about an underground forum marketplace. All the following case studies are only based on the derived supply chains. These case studies demonstrate the value and utility of extracting supply chains have in determining the origins of a specific crime or the use cases for products in our taxonomy. 

\subsection{\hf}

\subsubsection{Valuable accounts}

One common type of business in \hf, made apparent through supply chain link analysis, is selling social boosted accounts. This situation occurs when a user buys a \texttt{social-booster} to boost the ``follower'' or ``like'' count of a social media account, and then later sells the account for a premium because of the higher social status. Of the 589 (unattenuated) supply chain links extracted from the classified \hf\ data, 3\% were instances of social boosting and selling an account. A simple example that appeared in the \hf\ dataset was an instance where a user purchased a service which promised Twitter follows, and later sold a ``pre-made'' Twitter with 2k followers. Some of these groomed accounts were ``eWhore'' accounts (they are intended to appear to be owned by either an attractive man or woman), which we discovered are sold to romance scammers. This illuminated a connection between \texttt{social-booster} services and romance scammers which was not mentioned in prior work studying these scams~\cite{10.1007/978-3-319-20550-2_12}.

The value of an account is also dependent on the rarity of the handle. On \hf, these accounts are referred to as ``OG,'' which stands for ``original gangster'', and 32\% of our \texttt{account} links mention this term ~\cite{ogu}. Our supply chains depict these ``OG'' accounts exchanging hands, and an overwhelming 60\% of the links where ``OG'' is mentioned in the destination category come from an account source category. That being said, in order to obtain an ``OG'' account if an actor is not purchasing it directly, they must discover who the owner of the account is so that they can attempt to take over the account using methods such as phishing or SIM swapping attacks~\cite{krebsswap}. Furthermore, going from an ``OG'' username to personally identifiable information (PII) can happen through doxing~\cite{Snyder:2017}. In our taxonomy, we categorized doxing under hack-for-hire. There is an example link where the source category is an actor advertising a doxing service (hack-for-hire) and the purchaser of the service then sells a stolen ``OG'' account. 

\subsubsection{DDoS, botnets, and their crypter roots}

It is possible to understand the \texttt{botnet} supply chains through supply chain links in \hf. Many DDoS and botnet related criminal activities originate on \hf~\cite{krebsmirai}. For instance, the authors of the Mirai botnet, which launched a devastating attack against the Dyn DNS service, originally posted on \hf~\cite{krebsmirai}. Suppose there is a cybersecurity analyst which is interested in what purchases, or supply chain links, on \hf\ lead to someone selling a \texttt{botnet}. The categories which flow into \texttt{botnet} are \texttt{malware}, \texttt{proxy} and \texttt{account} and the categories which flow in \texttt{ddos-service} are \texttt{account}, \texttt{hosting}, \texttt{ddos-service}, \texttt{hack-for-hire}, \texttt{traffic}, \texttt{proxy}, \texttt{malware}, \texttt{video-game-service}, and \texttt{crypter}. Assuming further that the analyst is interested in the technical aspects of the \texttt{ddos-service}, they may be inclined to discover which crypters were purchased before a service is offered. These \texttt{crypter} to \texttt{ddos-service} chains are rare, and in fact this one type only makes up less than 1\% of all the found chains. Thus, for an analyst to stumble across this specific supply chain interaction on their own, they would need to find the needle in a haystack. With the help of our derived chains, we can see that the specific crypter works via Java drive-by download, which could lead an analyst to further investigate which systems are susceptible to this kind of exploit, infected with botnet malware obfuscated by the crypter, and carrying out DDoS attacks.

\subsection{Antichat}
The supply chains observed in Antichat can broadly be broken down into two main classes: \url{vk.com} - 412 links related (without attenuation) and other - 232 links related. The distinction between the classes is based on whether the service provided is related to \url{vk.com}. Contrary to common beliefs~\cite{slateTim}, it appears that much of the observed supply chains are centered around the Russian internal market rather than those outside of Russia. All of the following analysis is based on the extracted supply chains.

\begin{figure*}[!htb]
\centering
\includegraphics[width=0.95\linewidth]{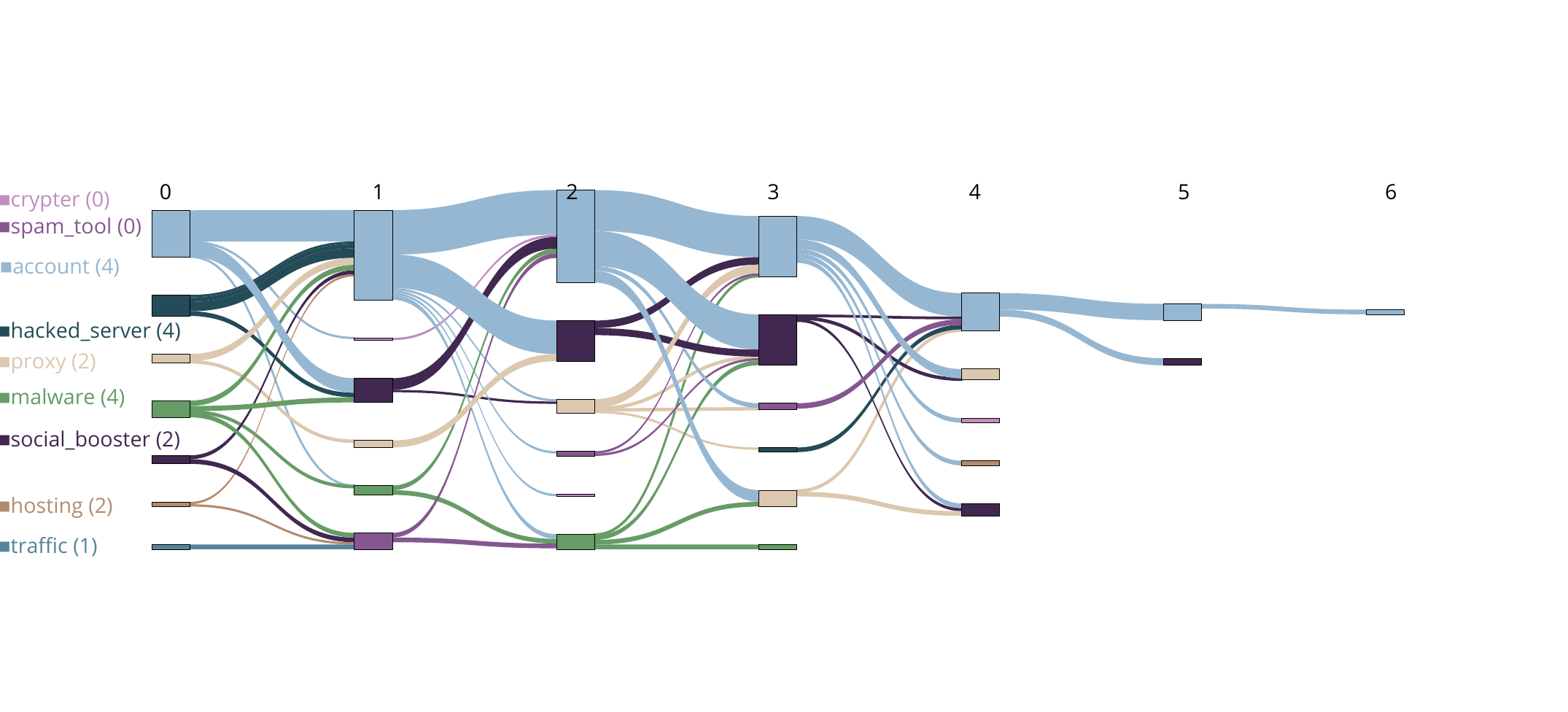}
\vspace{-0.5cm}
\caption{Supply chains found in Antichat, limited to links involving \url{vk.com}. Edges are colored according to source product category, and have widths determined by the number of users who purchased the source product and sold the destination product. Numbers at the top correspond to the level in the modified breadth-first search algorithm at which the node was discovered. The number of chains originating with each product category is denoted next to the product category names. Each user contributes to each link the amount determined by our method of attenuation.}
\label{fig:Antichat_Relevant_Alluvial_vk}
\includegraphics[width=0.95\linewidth]{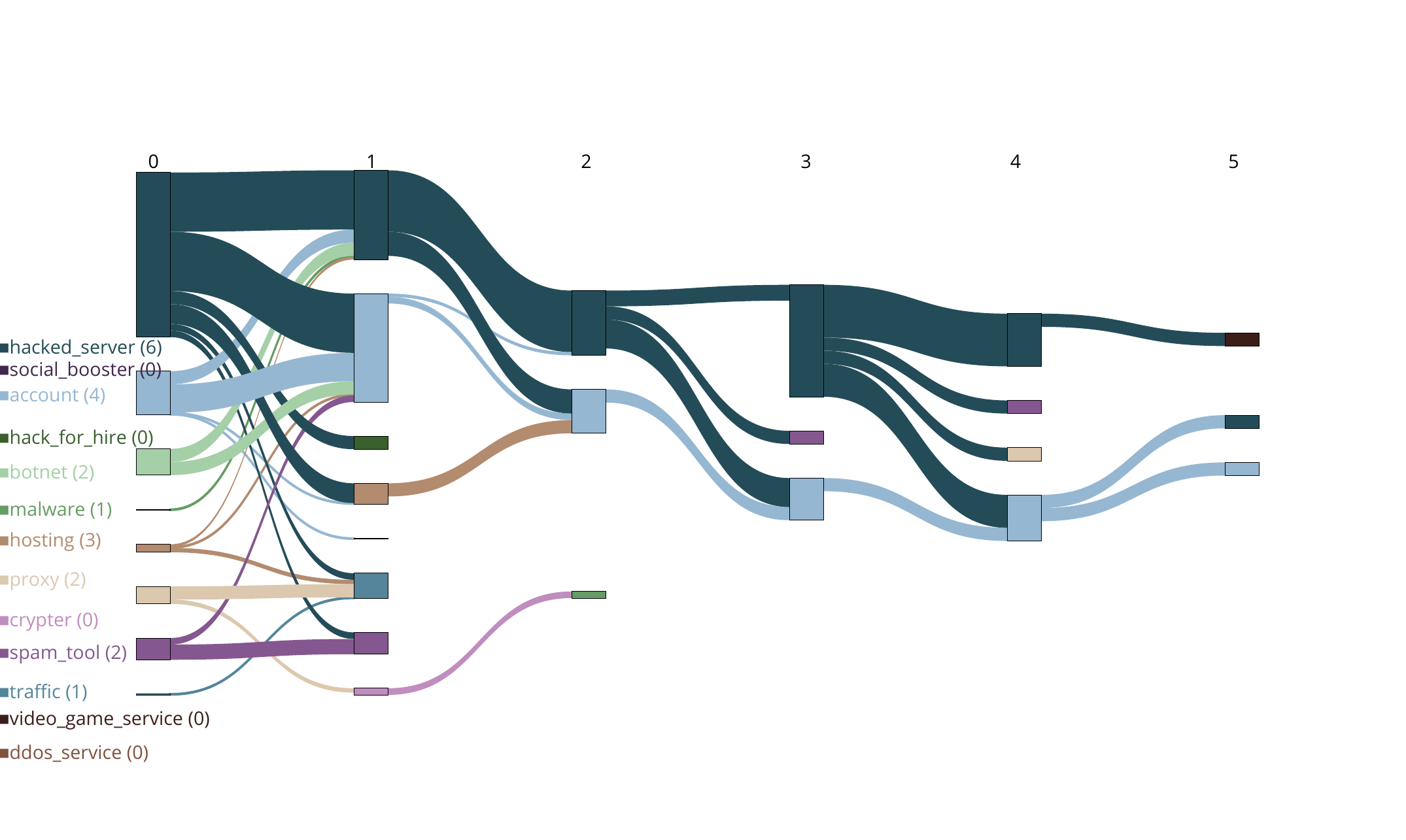}
\vspace{-0.5cm}
\caption{Supply chains found in Antichat, limited to links not involving \url{vk.com}, and validated as ``Related'' or ``Resell''. Edges are colored according to source product category, and have widths determined by number of users who purchased the source product and sold the destination product. Numbers at the top correspond to the level in the modified breadth-first search algorithm at which the node was discovered. Number of chains originating with each product category are denoted next to the product category names. Each user contributes to each link the amount determined by our method of attenuation.}
\label{fig:Antichat_Relevant_Alluvial_non_vk}
\vspace{-0.5cm}
\end{figure*}

\subsubsection{vk.com Relevant Supply chains}
Figure~\ref{fig:Antichat_Relevant_Alluvial_vk} shows the supply chains where either the seller or the buyer provided a service on the \url{vk.com} platform. It should be noted that in the context of vk, \texttt{proxy} means proxy accounts for spamming, \texttt{crypter} refers to a mechanism to hide the external link from vk's fraudulent websites detector, and \texttt{malware} refers to ways to steal accounts that are either root-kits or fake websites.  

The graphs showcase that, by far the, most bought and sold asset is an account. Those accounts are a crucial part of the infrastructure, and all of the services bought and sold end up sinking into accounts in at least one of the supply chain stages. 

First, it can also be seen that the accounts exchange hands quite a bit. We observed 5 different types of accounts: 
\begin{itemize}
    \item \textit{Clean (Chistie) accounts} -- accounts that are freshly acquired and previously belonged to real people. People buying those accounts were interested in either reselling the internal currency stored on the account, benefiting from the account's rating, using the account for later spamming, or providing some sort of social boosting, such as voting or inviting people to events. Additionally, criminals sometimes buy those accounts to use as a springboard to get access to other services with the same password. 
    \item \textit{Spammed (Prospam) accounts} -- accounts that were previously used to send out spam in the form of personal messages or messages on communal boards in different groups. Those accounts are likely about to be noticed by the anti-spam system, so criminals that are only interested in spam sell them at lower prices. The accounts can still be used for all of the other uses listed for Clean accounts.
    \item \textit{Invited (Proinvite) accounts} -- accounts that are used to invite people into various communities. Those are now considered very similar to spammed accounts, but at that time, \url{vk.com} treated them very differently in the detection phase, and criminals saw the difference between the two. Similarly to Spammed accounts, the accounts themselves can be used for all of the non-invite activities with low risks of getting caught. 
    \item \textit{Non-valid (Nevalid) accounts} -- accounts on which the password has either been changed or been incorrectly collected. Criminals can use those accounts to log into other services, e.g., their e-mail provider or other social networks.
    \item \textit{Auto-registered (Avtoreg) accounts} -- the accounts that were automatically registered and are used for the purposes of spam or social boosting. Those accounts look natural, have real human-like behavior, subscribe to groups, like messages, and occasionally make posts. 
\end{itemize}

Several types of these accounts exchanging hands are usually spammers and criminals interested in group-invite services, disposing of the accounts they can no longer use. At the same time, users providing social-booster services are buying more accounts than they sell, suggesting that the vast majority of the accounts used for such services are either getting banned and thus cannot be sold again, or are not getting caught at all and can be used forever.

The supply chains also suggest that there is quite a bit of reselling taking place. Out of 276 exchanges of hands, 136 were talking about the same product with either no or minor changes to the description of the product. This makes sense for services included in categories like \texttt{malware}, but are quite surprising for services like \texttt{proxy}, \texttt{social booster}, or \texttt{account}. There are two possible explanations as to why that might happen: 1) just as in any market, this market is ruled by supply and demand; 2) prior work has shown that users are more likely to buy from somebody reputable~\cite{motoyama2011analysis}. High rated users could be buying products from lesser rated users and reselling them for higher prices.

Criminals buying accounts consume \texttt{proxies}, \texttt{social boosters}, and \texttt{malware} to sell more \texttt{accounts}. Those supply chains suggest that the combination of the above provide an efficient way to harvest more \texttt{accounts}. At the same time, every \texttt{social booster} supply chain consumes a lot of \texttt{malware} and \texttt{accounts}, which suggests that there is an actual market driving \url{vk.com} account demand and the previously used accounts are getting aggressively banned. 

\subsubsection{non-vk.com-related supply chains}

The first thing that becomes apparent is that the largest chunk of the remaining supply chains is centered around \texttt{hacked servers} and their operation. The second largest group is \texttt{account}; however, those \texttt{accounts} are  mostly email accounts and not from \url{vk.com}. People acquire dedicated servers in order to brute-force either email accounts or other dedicated servers. Similarly, one can see that \texttt{malware} is being consumed and later feeds into \texttt{accounts} and \texttt{hacked servers}. One of those chains is a person who bought a bruter, and the other one is a person buying a remote exploit. Abuse-tolerant VPNs and proxies are being used in this infrastructure as well, and all rely on hacked dedicated servers. Some of the dedicated servers have clear indications that they were previously used for poker or spam, suggesting that reselling might be happening due to blacklisting.

\section{Discussion}
\label{discussion}

\paragraph{Performance}
Table \ref{table:performance_times} demonstrates the time required to derive the chains presented in the evaluation. The product classification, reply classification and BFS link generation was all run on machines with 64 vCPUs and 416 GBs of memory. 

\begin{table}[htb!]
\centering
\begin{tabular}{@{}lcc@{}}
\toprule
                       & Antichat    & \hf  \\ \midrule
Annotation             & $\sim$2720   & $\sim$2720  \\
Product classification & $\sim$32    & $\sim$27    \\
Reply classification   & $\sim$1     & $\sim$7     \\
BFS link generation    & \textless 1 & \textless 1 \\
Total (Minutes)        & 1393        & 1394        \\
Total (Hours)          & $\sim$23.21 & $\sim$23.23 \\ \bottomrule
\end{tabular}
\caption{Performance times for the entire process of supply-chain extraction. All values are in minutes unless stated otherwise. The annotation time is based on an expected ~10 seconds per post with a total of > 8000 posts per task (> 16000 per data set).}
\label{table:performance_times}
\end{table}


\
\paragraph{Limitations}

Despite both forums operating for almost a decade, we were only able to identify a few hundred supply chains across them. Yet, the analysis presented should be considered a lower bound estimation of the supply chains on the forums for the following reasons. First, the biggest reduction in the number of links considered was the choice of categories. Without limiting the links to those describing products that we are interested in (i.e., without filter posts classified as \texttt{other}), there were 1041 links in \hf\ and 9043 links in Antichat, which were reduced down to 692  (66\%) and 1179 (13\%), respectively. Note that not all of those links are actually related, but those numbers are indicative of the reduction in scale. 

Second, as we chose to prioritize the precision of the classifiers, our models were very conservative. Third, to re-emphasize the importance of precision, when choosing what counts as ``evidence'' of purchasing a product, our reply annotators were conservative, excluding mere indications of slight interest as ``evidence'' of purchasing in a reply. 

Finally, our supply chains were constructed from the public part of the forums that feature only a subset of interactions between criminals. Moreover, we only considered the replies users left under the corresponding selling post and not the reply under the member account.


We could have been less conservative and identified additional supply chains at the cost of increased false positives resulting in increased manual review time. 

We did not have access to the user profile pages where additional indications of buying could be found. However, these are public and could be collected and analyzed likely revealing additional chains. The challenge with using profile feedback is that it is often not attributed to a specific post.
Our methods are fairly agnostic only requiring the ability to discover selling and buying indicators. Thus, our methods could likely be extended to identify supply chains within other types of data sources such as online social networks and instant messaging cybercrime groups.  


\paragraph{Practical Usage}

Prior work demonstrated the problem with cross-domain prediction in underground forums~\cite{Durrett2017IdentifyingPI}. Therefore, we generate a separate model for each of the forums. In order to build a well-performing classifier for new forums, it would likely require labeling of around 6,000 -- 8,500 posts by a domain expert. These estimates are based on the learning curves included in the appendix (Figures~\ref{fig:product_learning_curves_russian} and \ref{fig:product_learning_curves_english}). From our experience, this takes about 2-3 person-days of effort for a domain expert.

It should be noted that even within a single forum content shift over time proves a barrier to product category classification. We performed tests to investigate a recommended re-annotation cycle. We compared the performance of the classifiers trained on 200 posts from different months and tested on the two last months of both Antichat and \hf' forums. For both of these forums we noticed that the accuracy was better the closer the training set was to the test set in terms of time. Antichat F1 scores went from about 0.15 at the farthest to about 0.35 at the closest, and \hf\ F1 scores went from about 0.05 at the farthest to about 0.4 at the closest. Since the training set size was 200, these scores are very low, but show that proximity in time between the training set and test set is important when classifying products on those forums. 

Finally, while classifying replies, we made use of the quote removal algorithm, which searched each reply for string quotes citing previous replies in the same thread. Implemented in a practical system, it would be more appropriate to use features specific to the forum to detect quotations, such as html tags.

\paragraph{Are there any other interesting supply chains?}
As was mentioned before, the supply chains extracted in this paper were mainly limited by the chosen categories while the product posts outside of them were classified as \texttt{other}. In the case of Antichat, our product taxonomy only described about a tenth of the links in the forum. Whilst skimming through the \texttt{other} links, we discovered a lot of related and interesting ones. For example, when \url{vk.com} introduced internal currency, many different users started buying services moving the currency in and out of \url{vk.com}. At the same time, we saw quite a few users buying and selling the development of \url{vk.com}-internal flash casinos. Similarly, when \url{vk.com} introduced a phone verification requirement for account registration, more chains were related to buying SMS-activations and phone \texttt{malware}. 

While those categories were not considered in this paper, our methods allow an analyst to capture those chains if needed by changing the category taxonomy.




\section{Conclusion}
\label{conclusion}

In this paper, we proposed, implemented, validated, and analyzed a set of methods that can identify underground cybercrime forum supply chains. Our approach is the first step toward automating the discovery of supply chains, which can significantly reduce the manual effort required to analyze these forums.  We have shown how those supply chains can be used to understand the collaboration in cybercriminal forums and help with providing insights into major security incidents such as the Mirai botnet attacks and Target data breach. 

Previous approaches struggled with the identification of supply chains within these forums, requiring tremendous amounts of analyst time and subject matter expertise. We show that it is possible to discover common connections between products within these large-scale forums based on our semi-supervised method. We show that our method can potentially be used to identify more effective methods of undermining attacks. More research is needed to fully automate to the discovery of supply chains.

{\normalsize \bibliographystyle{acm}
\bibliography{mybib}{}}
\appendices

\section{Learning Curves}
 \begin{figure}[!htb]
  \centering
  \begin{minipage}[b]{0.4\textwidth}
    \includegraphics[width=\textwidth]{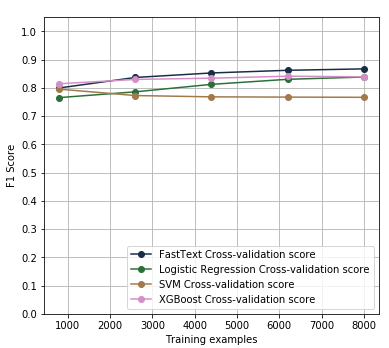}
    \caption{Reply Classification Learning Curves Russian}
    \label{fig:reply_learning_curves_russian}
  \end{minipage}
  \hfill
  \begin{minipage}[b]{0.4\textwidth}
    \includegraphics[width=\textwidth]{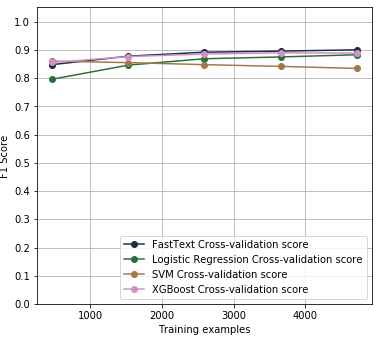}
    \caption{Reply Classification Learning Curves English}
    \label{fig:reply_learning_curves_english}
  \end{minipage}
\end{figure}

\begin{figure}[!ht]
 \centering
\begin{framed}
\begin{enumerate}
    \item \texttt{chicken proofreading service | fast, reliable | cheap you need homework done?}
    \item \texttt{snapbacks for sale!!! yo what's up hf i got 30+ snapbacks never worn before. }
    \item \texttt{[selling] lenovo erazer x700 decided to sell this bad boy since i use my macbook more than it.}
\end{enumerate}
\end{framed}
\caption{Examples \texttt{other} posts in \hf}
\label{fig:other_posts}
\end{figure}

\begin{figure}[!htb]
  \centering
  \begin{minipage}[b]{0.4\textwidth}
    \includegraphics[width=\textwidth]{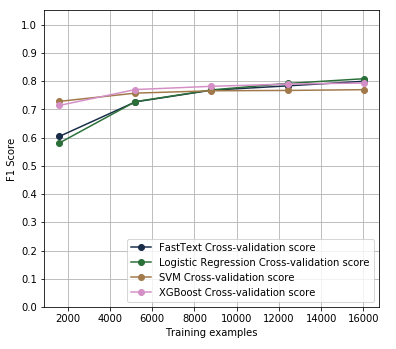}
    \caption{Product Classification Learning Curves Antichat (Russian)}
    \label{fig:product_learning_curves_russian}
  \end{minipage}
  \hfill
  \begin{minipage}[b]{0.4\textwidth}
    \includegraphics[width=\textwidth]{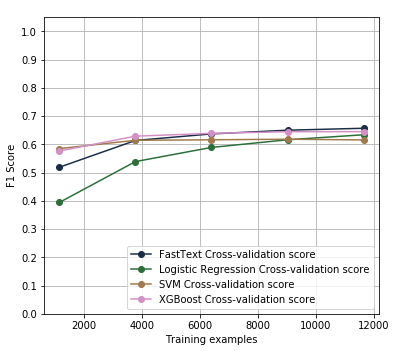}
    \caption{Product Classification Learning Curves Hackforums (English)}
    \label{fig:product_learning_curves_english}
  \end{minipage}
\end{figure}

\section{Confusion Matrix Heat Maps}
\begin{figure}[!htb]
  \centering
  \begin{minipage}[b]{0.4\textwidth}
    \includegraphics[width=\textwidth]{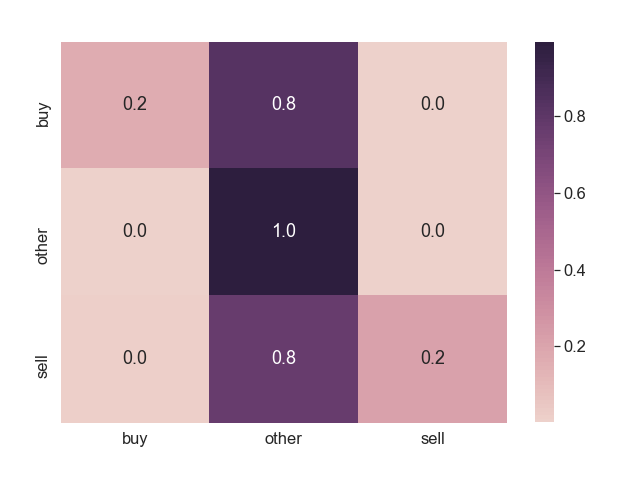}
    \caption{Reply Classification Confusion Matrix Antichat (Russian)}
    \label{fig:reply_cm_russian}
  \end{minipage}
  \hfill
  \begin{minipage}[b]{0.4\textwidth}
    \includegraphics[width=\textwidth]{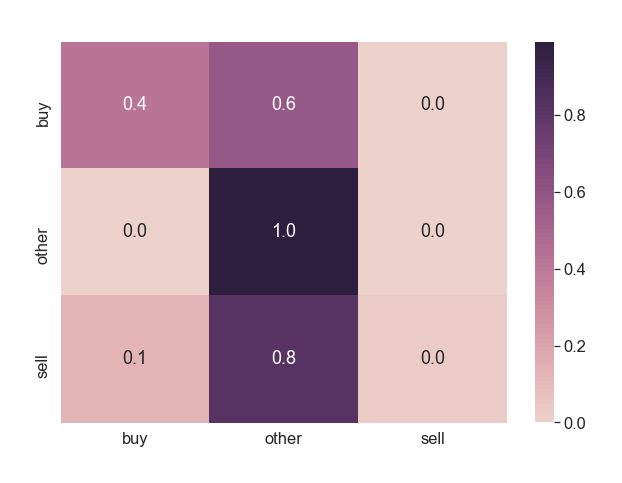}
    \caption{Reply Classification Confusion Matrix Hackforums (English)}
    \label{fig:reply_cm_english}
  \end{minipage}
\end{figure}

\begin{figure}[!htb]
  \centering
  \begin{minipage}[b]{0.4\textwidth}
    \includegraphics[width=\textwidth]{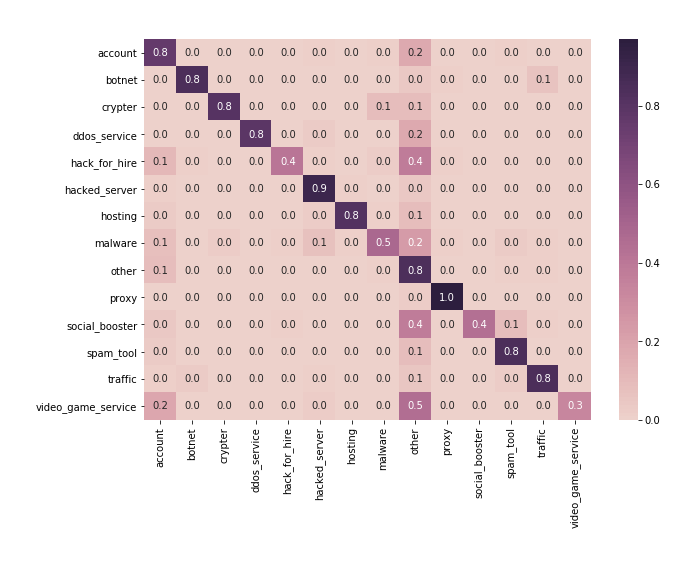}
    \caption{Product Classification Confusion Matrix Antichat (Russian)}
    \label{fig:product_cm_russian}
  \end{minipage}
  \hfill
  \begin{minipage}[b]{0.4\textwidth}
    \includegraphics[width=\textwidth]{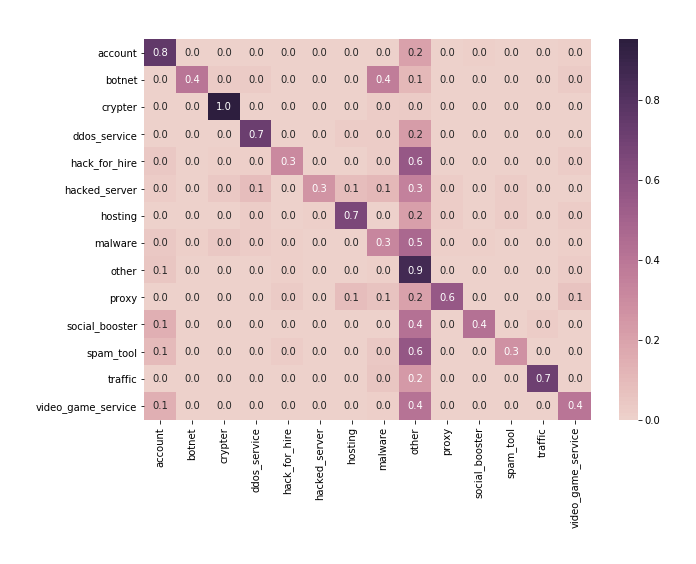}
    \caption{Product Classification Confusion Matrix Hackforums (English)}
    \label{fig:product_cm_english}
  \end{minipage}
\end{figure}

\end{document}